\documentclass[twocolumn,showpacs,amsmath,amssymb,prd]{revtex4}
\usepackage{graphicx}
\usepackage{graphicx}
\usepackage{dcolumn}
\usepackage{bm}
\usepackage{epsf}

\def\agt{\mathrel{\raise.3ex\hbox{$>$}\mkern-14mu\lower0.6ex\hbox{$\sim$}}}
\def\alt{\mathrel{\raise.3ex\hbox{$<$}\mkern-14mu\lower0.6ex\hbox{$\sim$}}}

\newcommand{\beq}{\begin{equation}}
\newcommand{\eeq}{\end{equation}}
\newcommand{\beqn}{\begin{eqnarray}}
\newcommand{\eeqn}{\end{eqnarray}}

\newcommand{\varep}{\varepsilon}

\begin{document}

\title{Magnetized hypermassive neutron star collapse: a central engine for
short \\ gamma-ray bursts}

\author{Masaru Shibata$^1$}

\author{Matthew D. Duez$^2$}
\altaffiliation{Current address:  Center for Radiophysics and Space Research,
Cornell University, Ithaca, NY 14853}

\author{Yuk Tung Liu$^2$}

\author{Stuart L. Shapiro$^2$}
\altaffiliation{Also at the Department of Astronomy and NCSA, University
of Illinois at Urbana-Champaign, Urbana, IL 61801}

\author{Branson C. Stephens$^2$}

\affiliation{$^1$Graduate School of Arts and Sciences, 
University of Tokyo, Komaba, Meguro, Tokyo 153-8902, Japan\\
$^2$Department of Physics, University of Illinois at Urbana-Champaign,
Urbana, IL 61801-3080}

\begin{abstract}
A hypermassive neutron star (HMNS) is a possible transient formed after 
the merger of a neutron star binary.  In the latest magnetohydrodynamic 
simulations in full general relativity, we find that a magnetized HMNS 
undergoes `delayed' collapse to a rotating black hole (BH) as
a result of angular momentum transport via magnetic braking and 
the magnetorotational instability.  The outcome is a BH surrounded by 
a massive, hot torus with a collimated magnetic field.   The torus 
accretes onto the BH at a quasi-steady accretion rate 
$\sim 10M_{\odot}$/s; the lifetime of the torus is $\sim 10$ ms. 
The torus has a temperature $\agt 10^{12}$ K, leading to copious 
($\nu\bar{\nu}$) thermal radiation.  Therefore, the collapse of an HMNS 
is a promising scenario for generating short-duration gamma-ray bursts
and an accompanying burst of gravitational waves and neutrinos.
\end{abstract}

\maketitle

Gamma-ray bursts (GRBs) are transient astrophysical phenomena that
emit large amounts of energy (typically $10^{51}$ ergs) in the gamma
ray band~\cite{GRB}.  The typical time variability is shorter than 10 
ms and the duration $t_{\rm dur}$ is $\sim 10$ ms--1000 s. These facts
suggest that the central engine of GRBs is a stellar-mass compact
object, and that the huge energy is supplied by converting
gravitational binding energy into radiation.  The popular theoretical
candidate for the central engine is a rotating stellar-mass black hole
(BH) surrounded by a massive, hot accretion torus (see~\cite{GRB} and
references therein).

Recent observations indicate that there are at least two classes of
GRBs: short-hard GRBs (hereafter SGRBs) with $t_{\rm dur} \sim 10$ ms--2 s
and long-soft GRBs with $t_{\rm dur}\sim 2$--1000 s.  For some long GRBs,
supernovae in spiral galaxies have been observed
coincidently~\cite{GRB-SN}, indicating that the central engine
(stellar-mass BH plus torus) for long GRBs is produced through stellar
core collapse of massive stars in the star forming region of spiral
galaxies. By contrast, associations between SGRBs and elliptical
galaxies have been reported~\cite{short}.  Since elliptical galaxies
have not produced massive stars in the past $\sim 10^{10}$ yrs, 
SGRBs are most probably not related to supernova stellar core collapse. 
In addition, recent observations of the afterglow of the SGRB
050709 rule out the presence of a supernova light curve and point to a
binary compact object merger as the most likely central
engine~\cite{fox2005}.

\begin{figure*}[t]
\begin{center}
\epsfxsize=2.1in
\leavevmode
\epsffile{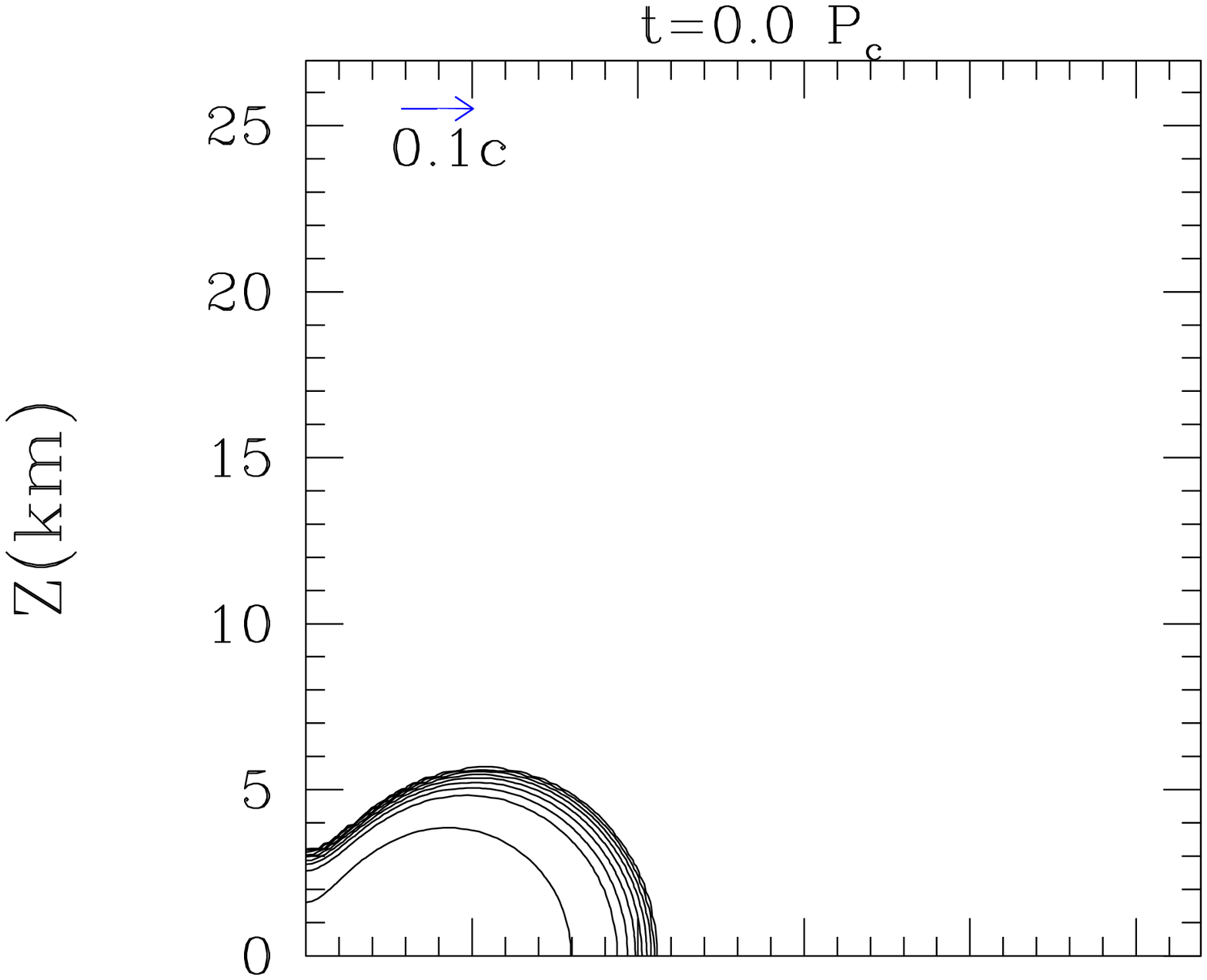}
\epsfxsize=2.1in
\leavevmode
\hspace{-1.8cm}\epsffile{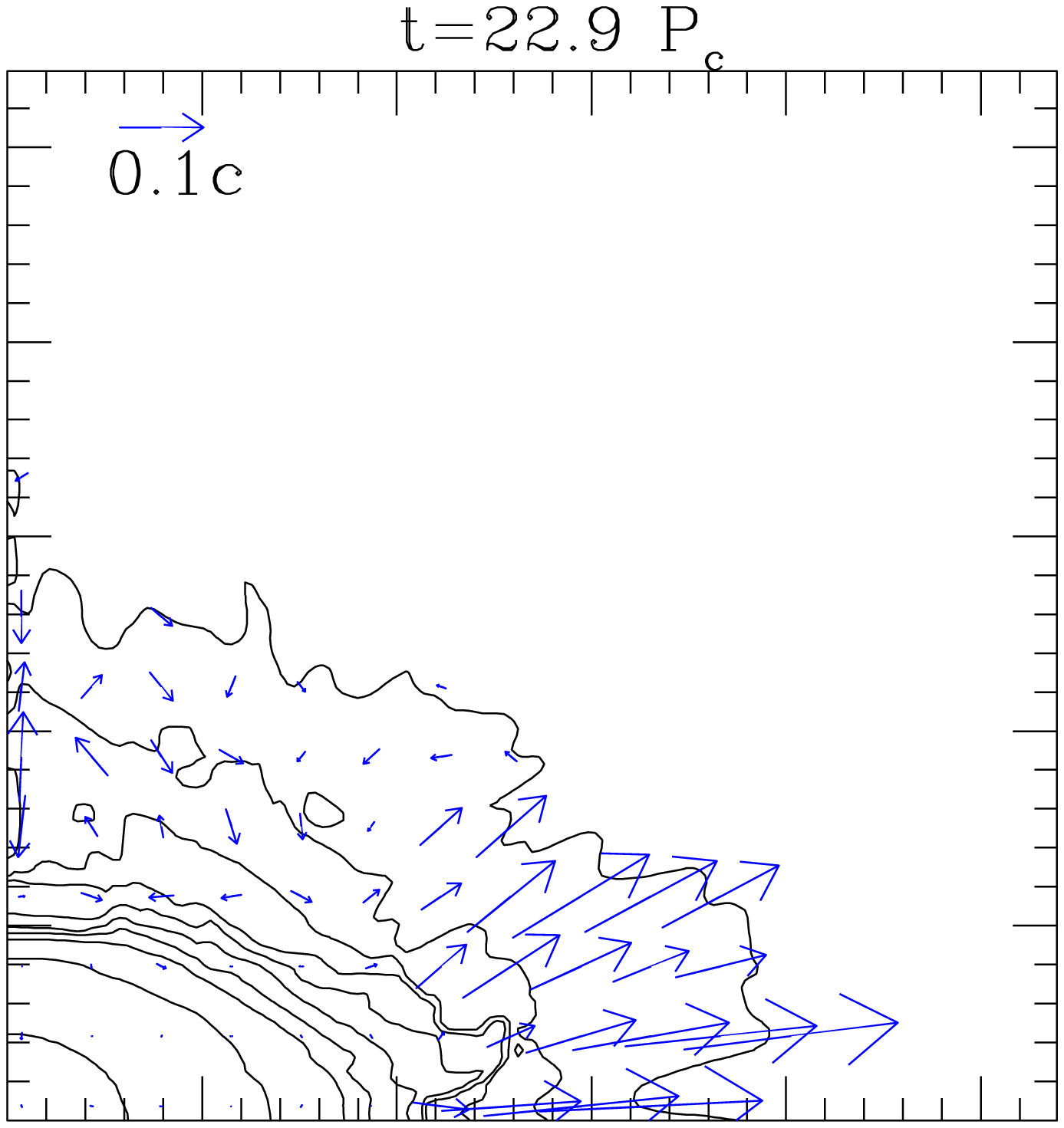}
\epsfxsize=2.1in
\leavevmode
\hspace{-1.8cm}\epsffile{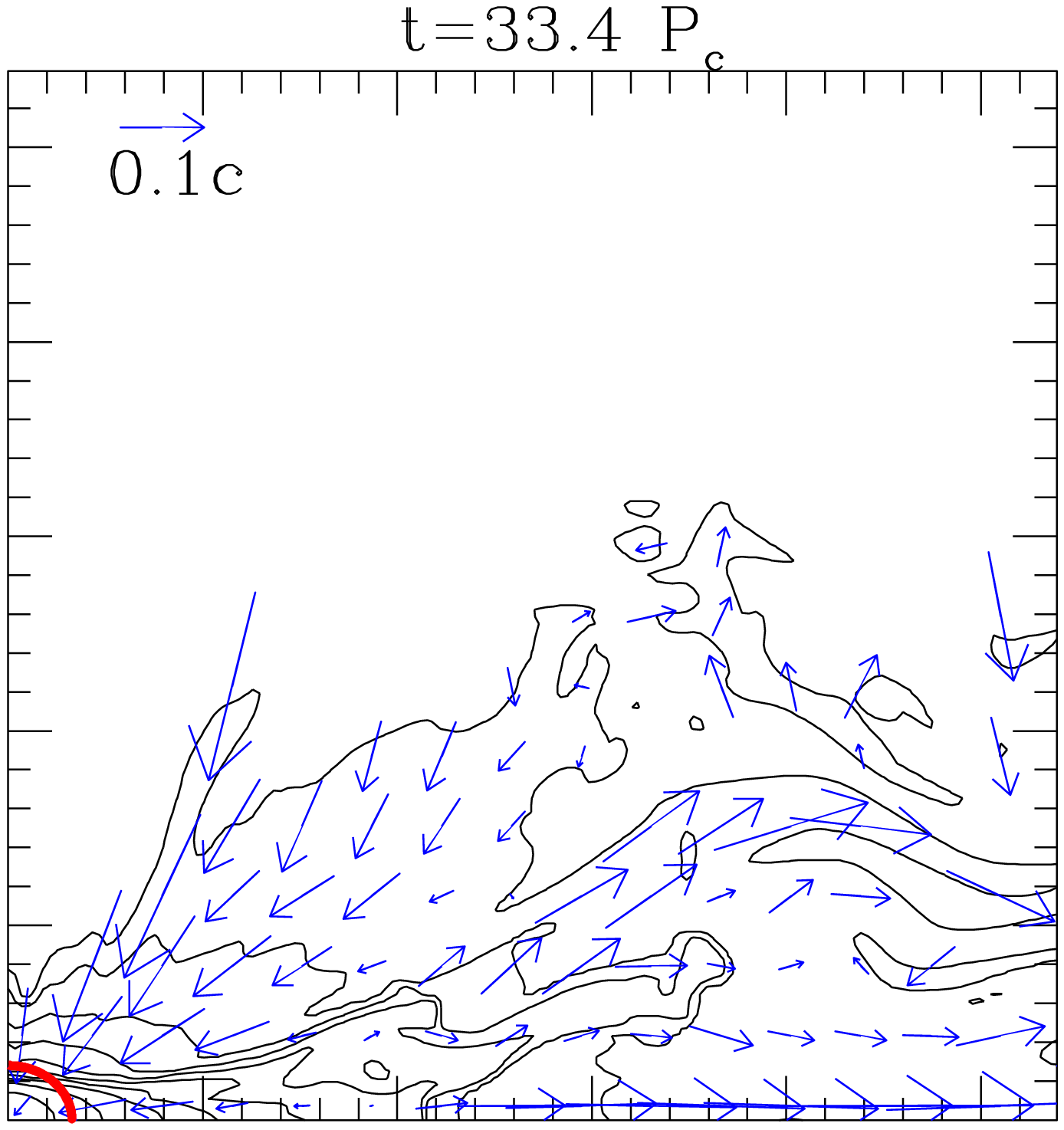}
\epsfxsize=2.1in
\leavevmode
\hspace{-1.8cm}\epsffile{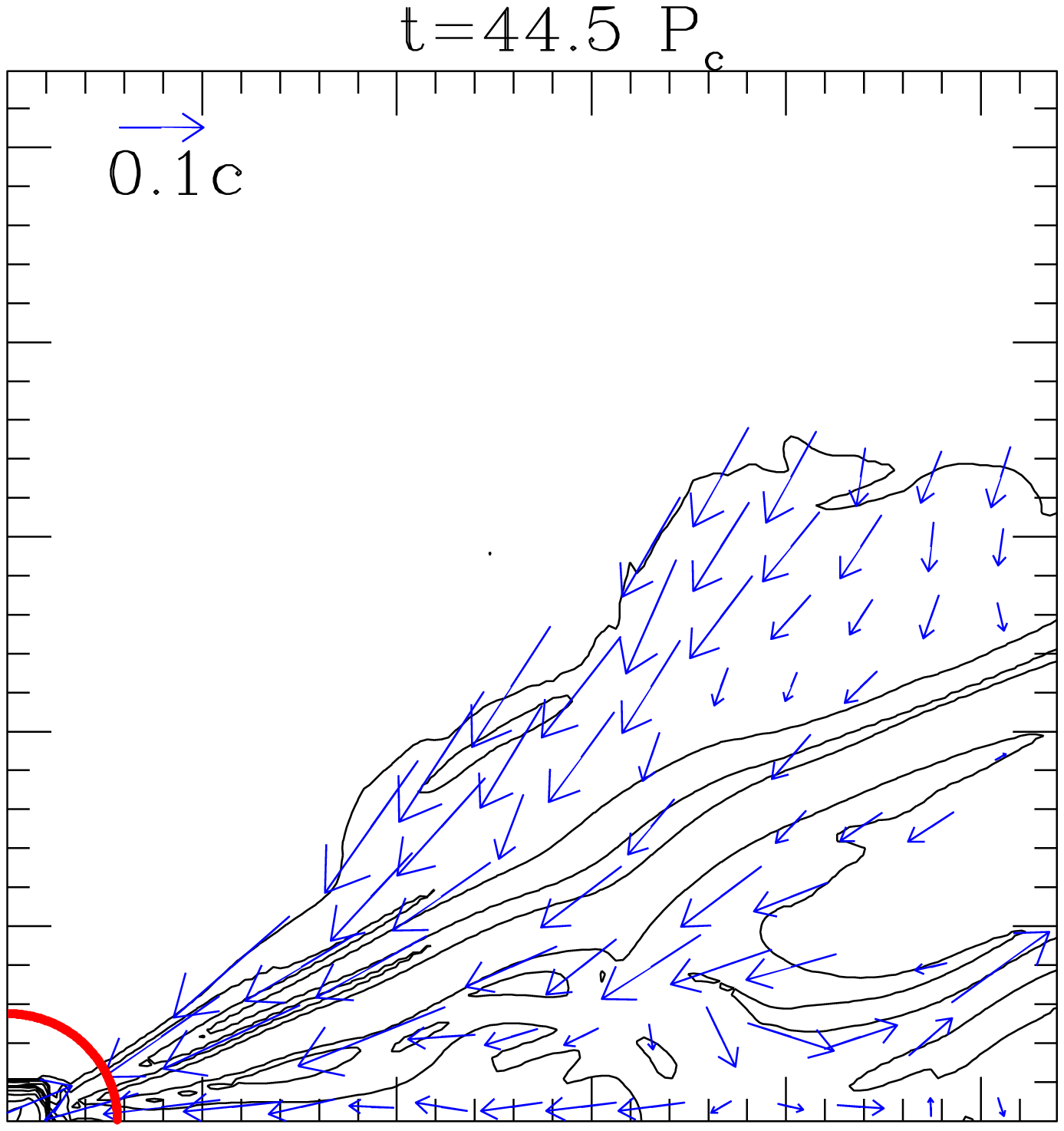}\\
\vspace{-1.7cm}
\epsfxsize=2.1in
\leavevmode
~\epsffile{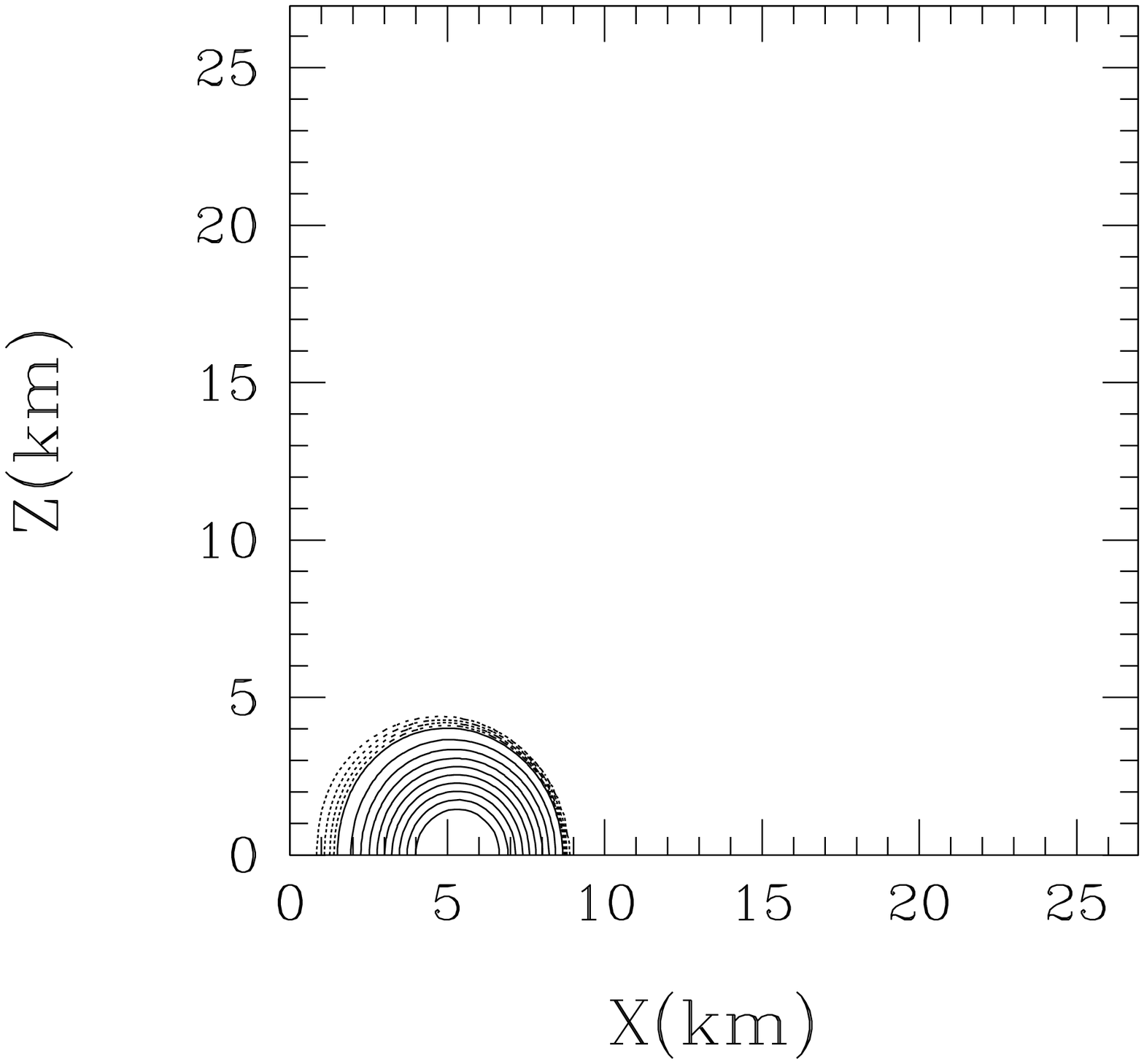}
\epsfxsize=2.1in
\leavevmode
\hspace{-1.8cm}\epsffile{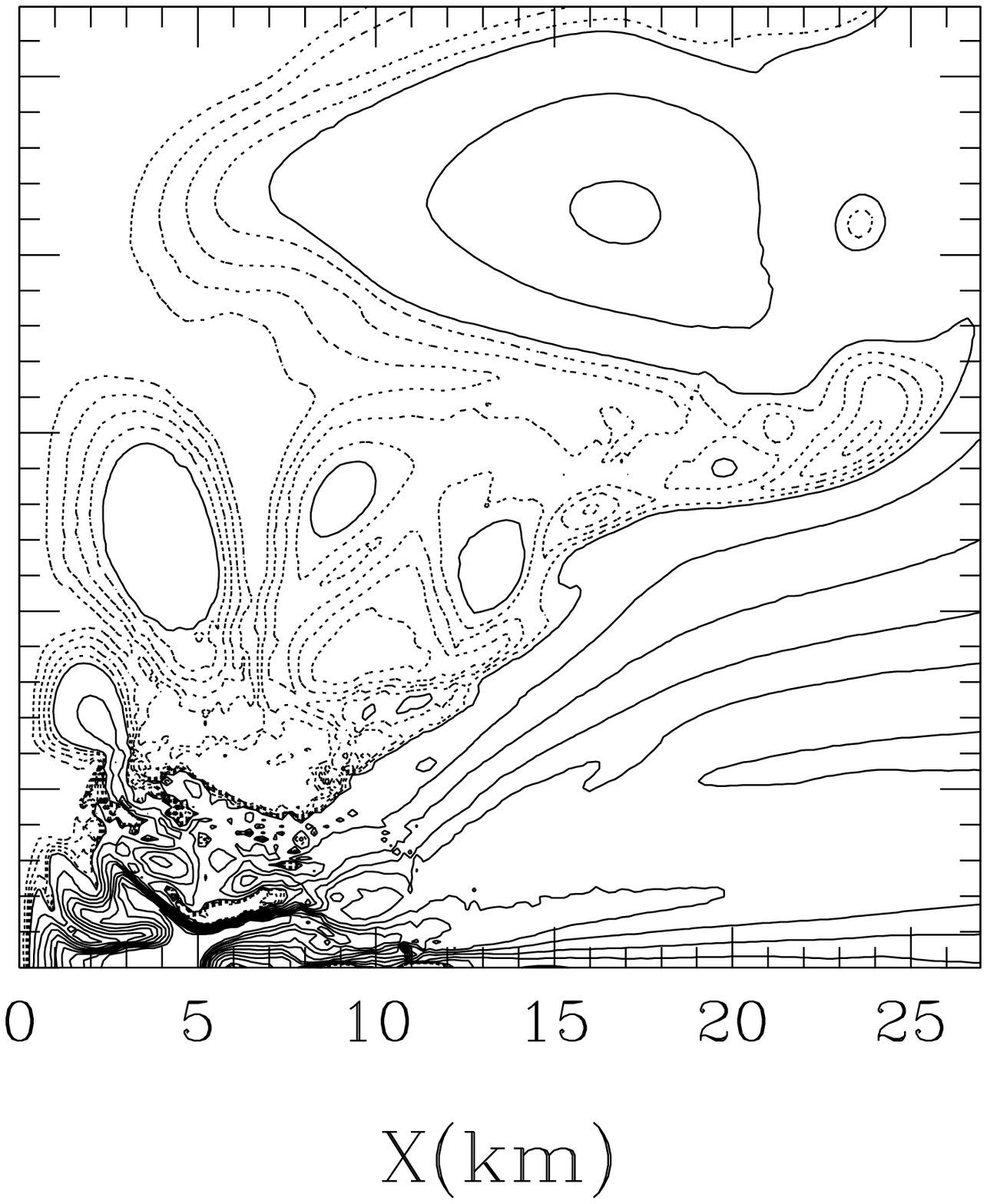}
\epsfxsize=2.1in
\leavevmode
\hspace{-1.8cm}\epsffile{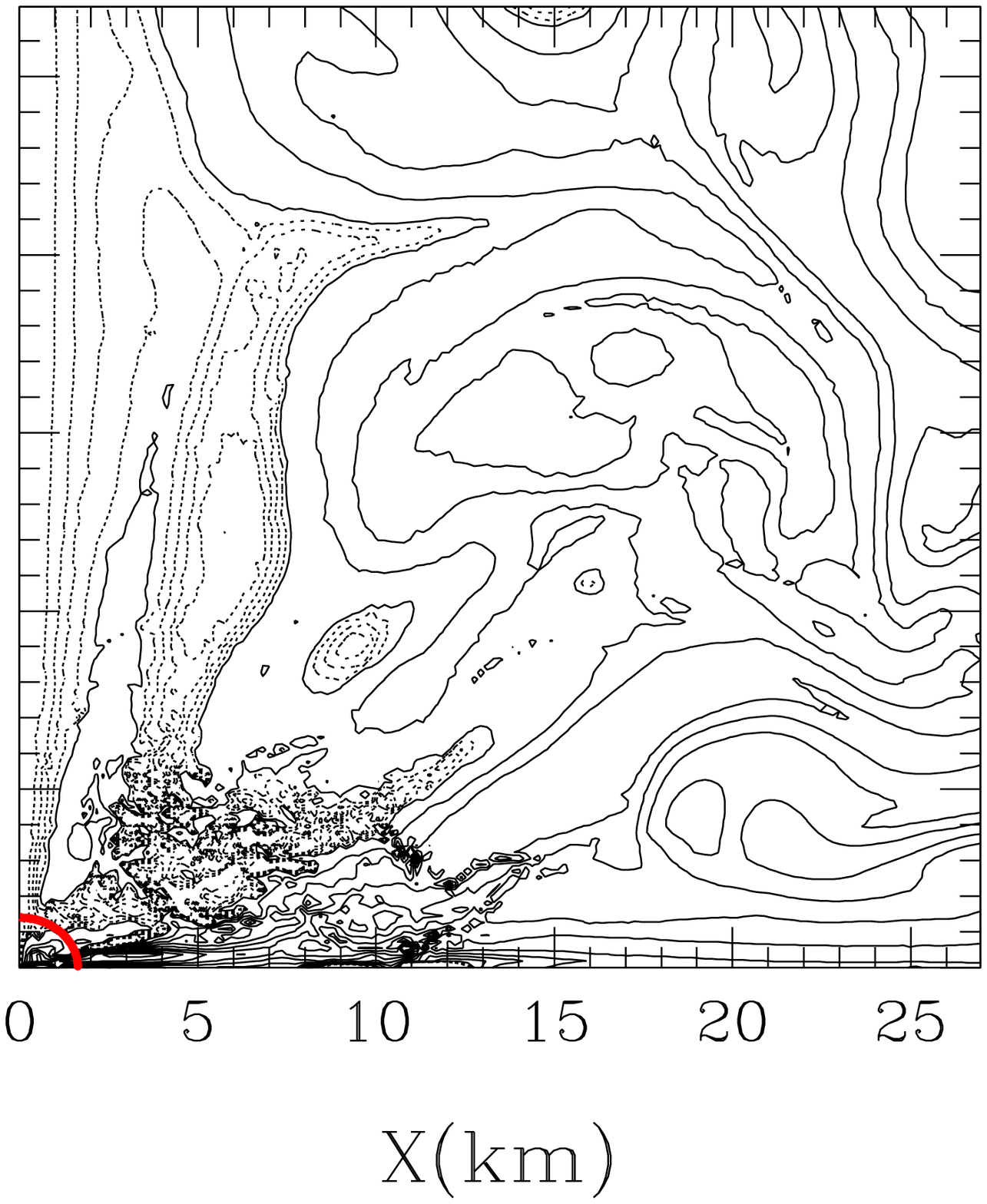}
\epsfxsize=2.1in
\leavevmode
\hspace{-1.8cm}\epsffile{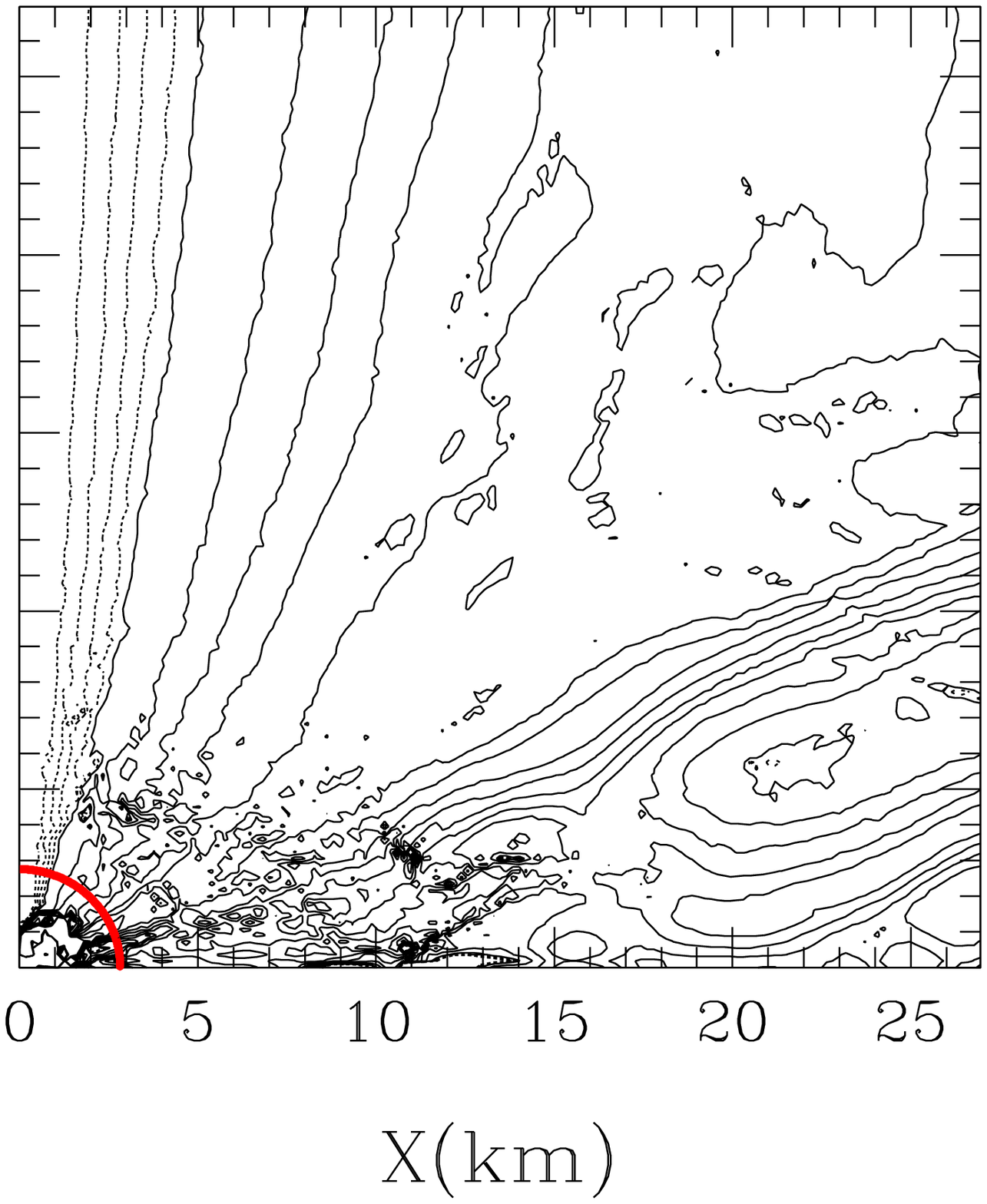}
\vspace{-6mm}
\caption{Upper four panels: Snapshots of the density contours
for $\rho$ (solid curves) and velocity vectors. The contours are
drawn for $\rho=10^{15}~{\rm g/cm^3} \times 10^{-0.4 i}~{\rm g/cm^3}~(i=0$--9).
In the last panel, a curve with $\rho=10^{11}~{\rm g/cm^3}$
is also drawn.  The (red) circle near the center in last two panels
denotes an apparent horizon.  The scale of the velocity is indicated in the
upper left corner.  The lower four panels denote the magnetic field
(contours of the toroidal component of the vector potential $A_{\varphi}$) at
the same times as the upper panels. The solid contour curves are drawn for
$A_{\varphi}= 0.8(1-0.1i)A_{\varphi,{\rm max,0}}~(i=0$--9) and
the dotted curves are for
$A_{\varphi}=0.08(1-0.2i) A_{\varphi,{\rm max,0}}~(i=1$--4).
Here,  $A_{\varphi}=A_{\varphi,{\rm max,0}}$ is the maximum value of
$A_{\varphi}$ at $t=0$.
\label{FIG1}}
\end{center}
\end{figure*}

The merger of binary neutron stars (BNSs) has been
proposed~\cite{GRB,GRB-BNS} as a candidate for SGRBs. According to
this scenario, after the merger, a stellar-mass BH is formed with an
ambient accretion torus of mass $\sim 1$--$10\%$ of the total. The
latest general relativistic hydrodynamic (GRHD) simulations (with no
magnetic fields) have shown that just after the merger of a BNS,
either a BH or a neutron star is formed \cite{STU,STU2}.  A BH forms
promptly if the total mass of the system, $M$, is larger than a
critical mass $M_{\rm thr}$. For mergers of nearly equal-mass BNSs
(the most likely case according to the data of observed binary
pulsars~\cite{Stairs}), far less than 1\% of the matter remains
outside the horizon, which is unfavorable for GRBs.  On the other
hand, for $M < M_{\rm thr}$, a hypermassive neutron star (HMNS) forms. 
Here, the mass is larger than the maximum allowed mass for rigidly
rotating neutron stars with an identical equation of state (EOS). An
HMNS is supported against collapse mainly by rapid and differential
rotation~\cite{BSS}. The temperature of an HMNS is high ($T \sim
10^{11}$ K) because of the heat generated by shocks during the merger
process. Although $T$ is high enough to produce a large amount of
neutrinos, neutrino-antineutrino ($\nu\bar\nu$) pair annihilation is
unlikely to generate a GRB fireball (consisting of relativistic
$e^+e^-$ pairs and photons).  This is because $\nu\bar\nu$
annihilation occurs primarily inside the HMNSs, and the available
energy is thus transferred to baryons (see e.g.,~\cite{RJ}). However,
HMNSs are transient objects and eventually collapse to BHs,
alleviating this `baryon loading problem.'

The value of $M_{\rm thr}$ depends crucially on the neutron star
EOS. Recent pulsar timing observations indicate~\cite{nice} the
existence of a neutron star of mass $2.1 \pm 0.2 M_{\odot}$ (one
$\sigma$ error). This measurement implies that the maximum mass of
spherical neutron stars, $M_{\rm sph}$, is larger than $\sim
2M_{\odot}$ and that stiff EOSs are favored. The latest GRHD
simulations with stiff EOSs like the one derived in~\cite{EOS}, in
which $M_{\rm sph} \approx 2$--$2.2 M_{\odot}$, indicate that $M_{\rm
thr}$ is $\approx 2.7$--$2.9M_{\odot}$. Thus, an HMNS is likely to be
formed after a merger of BNSs of canonical mass $\approx
2.6$--$2.8M_{\odot}$~\cite{Stairs} rather than a prompt collapse to a
BH.

The simulations in~\cite{STU2} also show that the HMNS remnants are
rapidly and differentially rotating and have triaxial shapes. These
HMNSs are secularly unstable since magnetic fields, viscosity, and/or
gravitational radiation will transport and/or dissipate angular
momentum and may trigger gravitational collapse. Recent numerical
simulations~\cite{STU2} suggest that gravitational wave emission may
trigger a collapse in $\sim 50$--100 ms for $M \agt 0.9M_{\rm thr}
\sim 2.4$--$2.6 M_{\odot}$ (this time scale will be longer for smaller
$M$).  In this case, the outcome will be a BH with a small disk ($\ll 0.01
M_{\odot}$), which is not a good candidate for the central engine of
SGRBs.

The other mechanisms which transport angular momentum are magnetic
braking~\cite{BSS,Shapiro} and the magnetorotational instability
(MRI)~\cite{MRI0,MRI}. These are likely to play a crucial role when
the magnetic fields in the HMNS are large enough. Magnetic
braking transports angular momentum on the Alfv\'en time
scale~\cite{BSS,Shapiro}, $\tau_A \sim R/v_A \sim 10^2 (B/10^{15}~{\rm
G})^{-1} $ $(R/15~{\rm km})^{-1/2}$ $(M/3M_{\odot})^{1/2}~{\rm ms}$, 
where $R$ is the radius of the HMNS. MRI occurs wherever
$\partial_{\varpi} \Omega < 0$~\cite{MRI}, where $\Omega$ is the
angular velocity and $\varpi$ is the cylindrical radius. This
instability grows exponentially with an e-folding time of $\tau_{\rm
MRI} =4 \left(\partial \Omega/\partial \ln \varpi
\right)^{-1}$~\cite{MRI}, independent of the field strength. For the
HMNS model considered in this paper, we find $\tau_{\rm MRI} \sim
1$~ms. When the MRI saturates, turbulence consisting of small-scale eddies often
develops, leading to angular momentum transport on a
timescale likely to be much longer than $\tau_{\rm MRI}$ \cite{MRI}. 

To study the effect of magnetic fields, we have performed general
relativistic magnetohydrodynamic (GRMHD) simulations for
differentially rotating HMNSs~\cite{DLSSS} using two new GRMHD
codes~\cite{DLSS2,SS}.  Here, we explore HMNS collapse further by
performing a simulation with the following hybrid EOS: $P=P_{\rm
cold}=K_1\rho^{\Gamma_1}$ for $\rho \leq \rho_{\rm nuc}$ and $P_{\rm
cold}=K_2\rho^{\Gamma_2}$ for $\rho \geq \rho_{\rm nuc}$. Here, $P$
and $\rho$ are the pressure and rest-mass density. We set
$\Gamma_1=1.3$, $\Gamma_2=2.75$, $K_1=5.16 \times 10^{14}$~cgs,
$K_2=K_1\rho_{\rm nuc}^{\Gamma_1-\Gamma_2}$, and $\rho_{\rm nuc}=1.8
\times 10^{14}~{\rm g/cm^3}$. With this EOS, the maximum gravitational
mass, $M$ (rest mass, $M_b$) is $2.01M_{\odot}~(2.32M_{\odot})$ for
spherical neutron stars and $2.27M_{\odot}~(2.60M_{\odot})$ for
rigidly rotating neutron stars. These are similar values to those in
realistic stiff EOSs~\cite{EOS}. We construct a differentially
rotating HMNS with the following characteristics: $M=2.65M_{\odot}$,
$M_b=2.96M_{\odot}$, maximum density $\rho_{\rm max}=9.0 \times
10^{14}~{\rm g/cm^3}$, angular momentum $J=0.82GM^2/c$, central
rotation period $P_c =0.202$~ms, ratio of polar to equatorial radius
$0.3$, and rotation period at the equatorial surface $5.4P_c$. The
rotation law is specified in the same way as in~\cite{DLSSS} with the
differential rotation parameter $\hat A=0.8$. This HMNS is similar to
that found in the BNS merger simulation of~\cite{STU2}.

For the simulation, a hybrid equation of state $P=P_{\rm
cold}+(\Gamma_{\rm th}-1)\rho(\varepsilon-\varepsilon_{\rm cold})$ is
used. Here, $\varepsilon$ is the specific internal energy, and $P_{\rm
cold}$ and $\varepsilon_{\rm cold}$ denote the cold part of $P$ and
$\varep$~\cite{SS}. The conversion efficiency of kinetic energy to
thermal energy in shocks is determined by $\Gamma_{\rm th}$, which we
set to 1.3 to conservatively account for shock heating. A seed
poloidal magnetic field is added to the HMNS by specifying the
$\varphi$-component of the vector potential as $A_{\varphi}=A_b
\varpi^2 {\rm max}[(P-P_{\rm cut}), 0]$ where $P_{\rm cut}$ is 0.04
times the maximum pressure and $A_b$ denotes a constant which
determines the initial strength of the magnetic fields. The value of
$A_b$ is chosen so that the maximum value of $C \equiv B^2/8\pi P$ at
$t=0$ is $3.42 \times 10^{-3}$. Here, $B^2/8\pi$ is the magnetic
pressure. This implies that the typical magnetic field strength is
$\sim 5 \times 10^{16}$ G.  Such a large value is chosen to save computational
time. Simulations with $1.9 \times 10^{-3} \alt C \alt 7.6 \times
10^{-3}$ indicate that a scaling relation approximately holds for
smaller seed fields from $t=0$ to BH formation via rescaling the time
as $t/t_A$. The simulation is performed with a uniform grid of size
$(N+1, N+1)$ for cylindrical coordinates $(\varpi, z)$, which cover
the region $[0, L]$ for each direction. Here, $L$ is chosen to be $5R$
($R \approx 2.75M = 10.8~{\rm km}$). The grid spacings are chosen as
$R/100$, $R/120$, and $R/150$ ($N=$500, 600 and 750), and approximate
convergence is confirmed.  Outside the HMNS, we add an atmosphere with
density $10^9~{\rm g/cm^3}$, which is necessary when employing
conservative schemes for the hydrodynamic equations.

\begin{figure}[tb]
\vspace{-2mm}
\begin{center}
\epsfxsize=2.9in
\leavevmode
\epsffile{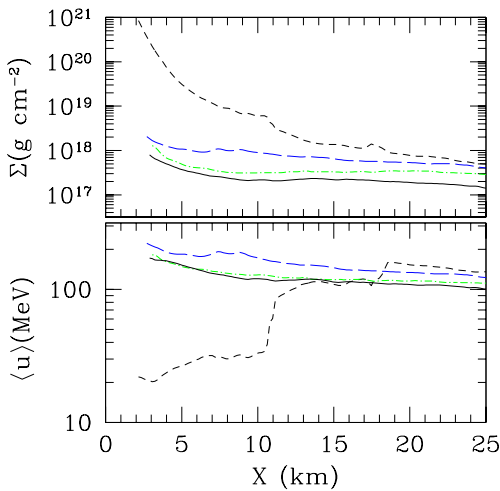}
\vspace{-10mm}
\caption{Evolution of the surface density and averaged thermal energy
per nucleon of the torus as functions of cylindrical radius at $t/P_c=33.4$ 
(dashed curves), 39.3 (long-dashed curves), 44.5 (dotted-dashed curves),
and 49.9 (solid curves). 
\label{FIG2}}
\end{center}
\end{figure}

In Figure~1, we show snapshots of the meridional density contours,
velocity vectors, and magnetic field lines for selected time
slices. Following an initial period of linear growth ($t \alt t_A
\approx 13 P_c$), the toroidal magnetic field begins to transport
angular momentum from the inner to the outer regions of the star
(magnetic braking), inducing quasistationary contraction of the
HMNS~\cite{DLSSS}.
At $t \sim t_A$, the growth of the toroidal magnetic field
saturates~\cite{Shapiro}. The subsequent evolution is dominated by
MRI~\cite{MRI}, which distorts the poloidal magnetic field lines and
leads to the formation of turbulent eddies on a scale much smaller
than $R$ (see the second lower panel of Fig.~1).  Because of the
turbulence, the matter located near the stellar surface is blown
outward. This expelled material, which is connected to the fluid in the
central region, further winds up the field lines, inducing additional 
magnetic braking. 

The star collapses at $t \simeq 33 P_c$, forming a BH composed of 
$\sim 85\%$ of the total rest mass (third panel of Fig.~1). 
Material with high enough specific angular momentum remains outside 
the newly formed BH and forms an accretion torus. However, the torus is 
secularly unstable, since magnetically-induced turbulence transports angular 
momentum outward.  The growth of the BH by quasistationary accretion 
is followed by employing an excision algorithm~\cite{excision}. 
The accretion rate $\dot M$ gradually decreases and eventually settles 
down to $\dot M \sim 10M_{\odot}/$s. At $t \sim 50P_c$, the 
rest mass of the torus is $\sim 0.05M_{\odot}$, and the total accretion time 
is thus $\approx 20P_c + 0.05M_{\odot}/\dot M \sim 10~{\rm ms}$. 
Note also that a collimated magnetic field has formed along the rotation 
axis (the rightmost lower panel of Fig.~1).

To clarify the properties of the torus, we calculate the surface 
density $\Sigma$ and the vertically averaged thermal energy per nucleon,
$\langle u \rangle$ (see Fig.~2).  The local thermal energy per nucleon 
is given by $u=m_N\varepsilon_{\rm th}$, where the thermal
part of the specific internal energy is
$\varepsilon_{\rm th}\equiv \varepsilon - \varep_{\rm cold}$, and
where $m_N$ is the mass of a nucleon.  (We assume that the torus is 
composed of free nucleons.) Thus we have
\beqn
&&\Sigma(\varpi) = \int_{z \geq 0} \rho u^t \sqrt{-g} dz,\\
&&\langle u \rangle (\varpi)=\frac{m_N}{\Sigma(\varpi)}
\int_{z \geq 0} \rho u^t \sqrt{-g} \varepsilon_{\rm th} dz, 
\eeqn
where $g$ and $u^t$ denote the determinant of the spacetime metric and 
the time component of the four velocity.  The integrals are carried out 
along lines of $\varpi$=constant. Note that $\varep_{\rm th}$ is 
zero at $t=0$ inside the HMNS and subsequently grows due to shock
heating. The typical thermal energy per nucleon is 
$u\approx 94 ({\varep_{\rm th}/0.1c^2})$~MeV/nucleon, 
or equivalently, $T \approx 1.1 \times 10^{12}
(\varep_{\rm th}/0.1c^2)$ K. 

Because of its high temperature and density, the torus radiates
strongly in thermal neutrinos~\cite{PWF,MPN}. However, the opacity
inside the torus (considering only neutrino absorption and scattering
interactions with nucleons) is $\kappa \sim 7\times 10^{-15}
T_{12}^2$~${\rm cm}^2$~${\rm g}^{-1}$, so that the neutrinos are
effectively trapped~\cite{MPN}. Here, $T_{12} = T/10^{12}~{\rm K}$.
Since the torus is optically thick, the neutrino luminosity may be
estimated in the diffusion limit~\cite{ST} as $L_{\nu} \sim \pi R^2
F$, where $R$ is the typical radius of the emission zone, the flux is
$F \sim (c/3)(7N_{\nu}/4)(\sigma T^4/\tau)$ (where $\sigma$ is the
Stefan-Boltzmann constant and $N_{\nu}$ is the number of neutrino
species, taken as 3), and the neutrino optical depth is \mbox{$\tau
\sim \kappa \Sigma$}.  Then $L_{\nu} \sim 2\times 10^{53}~{\rm erg/s}
(R/10~{\rm km})^2 T_{12}^2 \Sigma_{18}^{-1}$, which is comparable to
the neutrino Eddington luminosity~\cite{MPN}.  Here,
$\Sigma_{18}=\Sigma/10^{18}~{\rm g}~{\rm cm}^{-2}$.  Because of the
geometry of the torus, pair annihilation will be most efficient near 
the $z$-axis.  Furthermore, the surface density along the $z$-axis
($\varpi = 0$) outside the apparent horizon is $\Sigma_{18} \sim 0.01$
for $t \agt 40 P_c$, which is much smaller than the surface density of
the torus.  In fact, the total mass contained in a cylinder of radius
$\varpi \sim M(\sim 5 {\rm km})$ is $\sim 10^{-6} M_{\odot}$, which is
likely small enough to allow the formation of a relativistic
fireball~\cite{GRB}.

Our numerical results suggest the presence of a hot, hyperaccreting
torus which is optically thick to neutrinos.  A model for the neutrino
emission in a similar flow environment with comparable $L_{\nu}$ (a
`neutrino dominated accretion flow') is provided by Di Matteo et
al.~\cite{MPN}.  According to this model, the luminosity due to
$\nu\bar\nu$ annihilation is $L_{\nu\bar\nu}\sim 10^{50}~{\rm
ergs/s}$~\cite{MPN}. Aloy et al.~\cite{AJM} simulate the propagation
of jets powered by energy input along the rotation axis (as would be
supplied by the $\nu\bar\nu$ annihilation).  They find that if the
half-opening angle of the energy injection region is moderately small
($\alt 45^{\circ}$) and the baryon density around the BH is
sufficiently low, jets with the Lorentz factors in the hundreds can be
produced given an energy input $L_{\nu\bar\nu}\agt 10^{48}~{\rm
ergs/s}$ lasting $\sim 100$ ms.  They also show that the duration of
SGRBs may be $\sim 10$ times longer than the duration of the
energy input because of the differing propagation speeds of the jet
head and tail.  Our numerical results, along with the accretion flow
and jet propagation models of~\cite{MPN,AJM}, thus suggest that
magnetized HMNS collapse is a promising candidate for the central
engine of SGRBs. Since the lifetime of the torus is $\sim 10$ ms
in our simulation, the total energy of the $\nu\bar\nu$ annihilation
($E_{\nu\bar\nu} \sim 10^{48}~{\rm ergs}$) may be sufficient to power
SGRBs as long as the emission is somewhat beamed (and beaming is
probably encouraged by the fat geometrical structure of the
torus~\cite{AJM}).

Alternatively, a relativistic outflow could also be powered by MHD
effects~\cite{GRB}. Though the GRMHD equations are solved
self-consistently in our simulation, we do not find evidence for
strong MHD outflows.  This may be a consequence of our initial
magnetic field configuration or our neglecting neutrino pressure, and
requires further study.  However, simulations of magnetized accretion
tori in fixed Kerr spacetime~\cite{MG} have found outgoing
electromagnetic energy due to the Blandford-Znajek (BZ)
effect~\cite{BZ}.  The BZ luminosity~\cite{MG} is estimated as
$L_{\rm BZ} \sim 10^{53} a^2 (B /10^{16}~{\rm G})^2
(M / 2.8 M_{\odot})^2~{\rm erg/s}$, where  
$a$ is the nondimensional BH spin parameter and $B$ is the
typical magnetic field strength.  Assuming a reasonable conversion
efficiency from the Poynting flux to the kinetic energy of
the fireball and then to gamma-ray energy, energy fluxes of this
magnitude are sufficient for forming SGRBs.

Finally, this model predicts that SGRBs should accompany a burst
of gravitational radiation and neutrino emission from the HMNS delayed
collapse. We plan to study this gravitational radiation in a future
paper~\cite{big}. 

{\em Acknowledgments}: MS thanks K. Ioka and R. Takahashi for helpful
comments. Numerical computations were performed on the FACOM VPP5000
at ADAC at NAOJ, on the NEC SX6 at ISAS at JAXA, and at the NCSA at
UIUC.  This work was supported in part by Japanese Monbukagakusho
Grants (Nos.\ 17030004 and 17540232) and NSF Grants PHY-0205155 and
PHY-0345151, NASA Grants NNG04GK54G and NNG046N90H at UIUC.

\end{document}